\documentclass[sigplan,screen,nonacm]{acmart}

\usepackage[utf8]{inputenc}
\usepackage{array}
\usepackage{hyperref}

\begin{document}

\title{StableSims}
\subtitle{Optimizing MakerDAO Liquidations 2.0 Incentives via Agent-Based Modeling}
\author{Andrew Kirillov}
\affiliation{%
    \institution{Blockchain at Berkeley}
    \city{Berkeley}
    \state{California}
    \country{USA}
}
\email{akirillo@berkeley.edu}

\author{Sehyun Chung}
\affiliation{%
    \institution{Blockchain at Berkeley}
    \city{Berkeley}
    \state{California}
    \country{USA}
}
\email{sehyun@berkeley.edu}

\date{January 2022}

\begin{abstract}
    The StableSims project set out to determine optimal parameters for the new auction mechanism, Liquidations 2.0, used by MakerDAO, a protocol built on Ethereum offering a decentralized, collateralized stablecoin called Dai. We developed an agent-based simulation that emulates both the Maker protocol smart contract logic, and how profit-motivated agents (“keepers”) will act in the real world when faced with decisions such as liquidating “vaults” (collateralized debt positions) and bidding on collateral auctions. This research focuses on the incentive structure introduced in Liquidations 2.0, which implements both a constant fee (\texttt{tip}) and a fee proportional to vault size (\texttt{chip}) paid to keepers that liquidate vaults or restart stale collateral auctions. We sought to minimize the amount paid in incentives while maximizing the speed with which undercollateralized vaults were liquidated. Our findings indicate that it is more cost-effective to increase the constant fee, as opposed to the proportional fee, in order to decrease the time it takes for keepers to liquidate vaults.
\end{abstract}

\maketitle

\section{Introduction}
This paper assumes that the reader is generally familiar with MakerDAO’s Dai stablecoin system, and so we will not endeavor into an overview of the protocol. For this, please read their whitepaper \cite{Maker21a}. With that said, it’s worth exploring the core differences between Liquidations 2.0 and its predecessor, Liquidations 1.2, and the motivation for this change.

\subsection{Liquidations 1.2 vs Liquidations 2.0}
The primary difference between Liquidations 2.0 and the previous implementation, Liquidations 1.2, is the underlying auction mechanism used to sell off liquidated collateral and pay off the associated Dai loan.

In Liquidations 1.2, a “classic,” or English, auction was used. English auctions run for a fixed amount of time, and over the course of the auction, bids of increasing value are placed until the auction duration has expired, at which point the “lot” (item up for auction, in our case vault collateral) goes to the highest bidder.

There are two primary issues that motivated the redesigning of the auction mechanism, laid out in the Liquidations 2.0 discussion post on the MakerDAO forum \cite{Maker20a}:

\begin{enumerate}
    \item Reliance on DAI liquidity. Under Liquidations 1.2, keepers must have sufficient DAI capital to place a bid, and mus lock that bid up for multiple blocks.
    \item Likelihood of auction settlement far from market price. Under Liquidations 1.2, a near-zero initial bid can be placed, and could go uncontested due to network congestion, or liquidity constraints implied by the properties above. This is exactly what happened during MakerDAO's "Black Thursday" event \cite{Maker20b}.
\end{enumerate}

To tackle these issues, Liquidations 2.0 uses a Dutch auction system. In a Dutch auction, there is a high initial bid that gradually decreases until someone agrees to “take” the lot at that current bid price.

This mitigates the first issue via instant auction settlement: an auction settles within the same block where a keeper takes the lot. This allows for flash-loaning of auction bids, removing capital requirements for keepers.

The second issue is addressed by relying on MakerDAO's Oracle Module to set an auction's initial bid price, and prevent the price from dropping too far by requiring that the auction be restarted by a keeper under either of two conditions:

\begin{enumerate}
    \item The auction has gone on for too long
    \item The bid price of the auction has dropped by a sufficient percentage
\end{enumerate}

Both of these conditions are parameterized by constants in the auction contract for the given collateral type. When an auction is restarted, a new initial bid price is pulled from the Oracle Module. Thus, it's impossible for the auction to settle too far from the collateral's market price.

More information about the Liquidations 2.0 system can be found on its documentation page \cite{Maker21b}, and in its Maker Improvement Proposal post on the forum \cite{Maker21c}.

\subsection{\texttt{chip} and \texttt{tip}}
Some incredible research has been done by Gauntlet on Liquidations 2.0 \cite{Gauntlet21a} - also using agent-based modeling - to recommend optimal settings for the vast majority of parameters in the new auction system, including those responsible for determining the auction’s initial price, price curve, and duration. However, their research omitted any insight into the newly introduced \texttt{chip} and \texttt{tip} parameters, which are used to incentivize the liquidation of undercollateralized vaults and the restarting of stale collateral auctions.

The \texttt{chip} parameter defines a proportion of the liquidated vault’s Dai loan to award to the liquidating or restarting keeper. For example, if a vault with 100 DAI drawn as debt is liquidated, and the \texttt{chip} value for the associated collateral type is 0.05, then 5 DAI will be minted for the keeper who liquidated the vault.

The \texttt{tip} parameter defines a flat fee to be paid to the liquidating or restarting keeper, regardless of the size of the vault they’re liquidating. It depends only on the collateral type associated with the vault. For example, if \texttt{tip} is set to 100 for ETH-A vaults, then a keeper that liquidates any ETH-A vault will be rewarded with 100 DAI.

The total incentive a keeper receives for liquidating an undercollateralized vault is given by:
$$\texttt{chip} \cdot \texttt{vault\_debt} + \texttt{tip}$$

In exploring optimal settings for \texttt{chip} and \texttt{tip}, the tradeoff we considered was minimizing these values, and thus the amount of incentive paid out by the protocol, while maximizing the speed with which undercollateralized vaults were liquidated. Sufficiently incentivized keepers would not need to wait for more favorable market conditions such as lower transaction fees or more exchange liquidity in order to trigger liquidations.

\section{Methodology}
\begin{figure}[t]
  \centering
  \includegraphics[width=\linewidth]{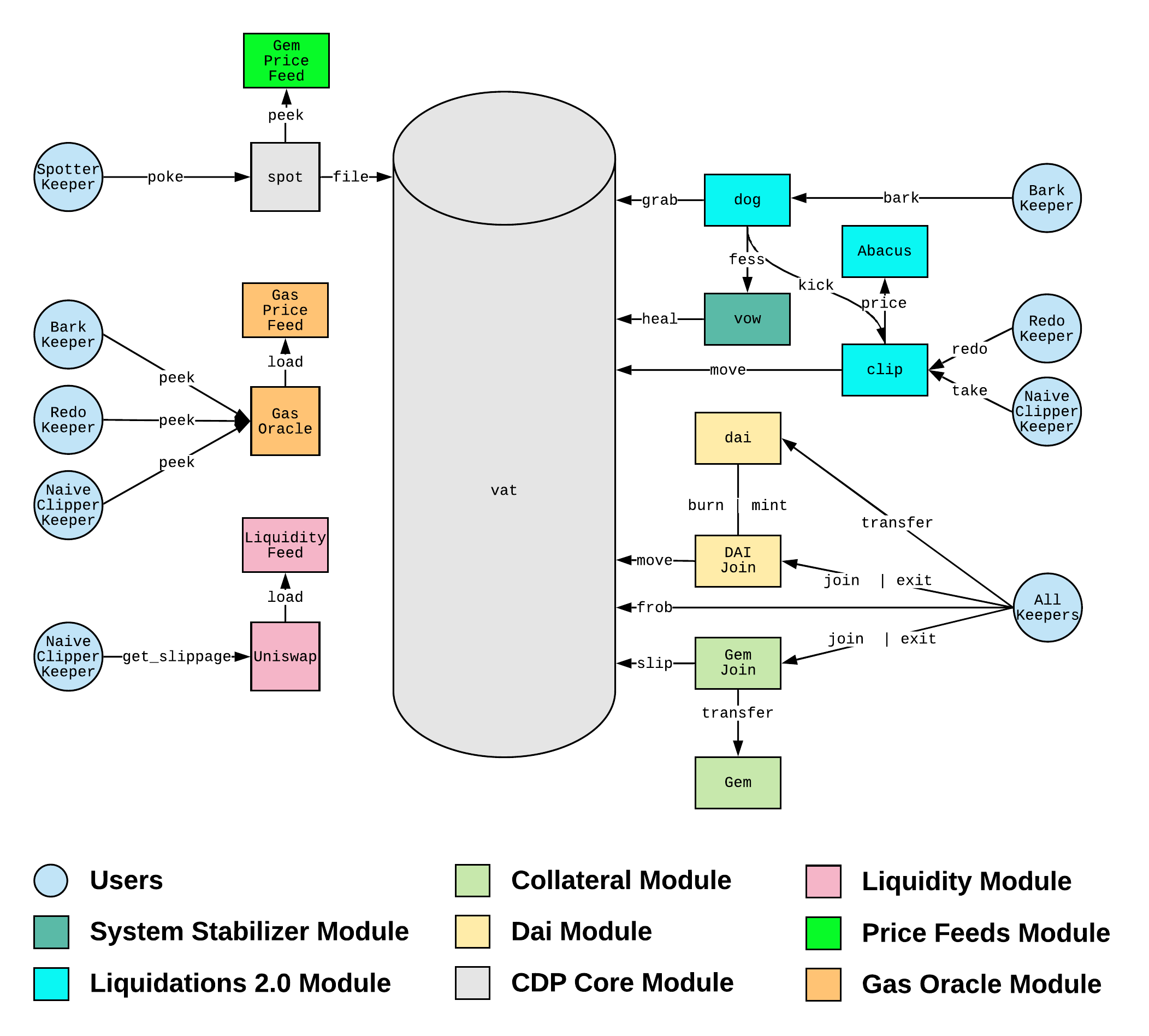}
  \caption{StableSims system diagram}
  \label{system}
\end{figure}

To create the simulation, we relied heavily on the open source code that the Maker protocol runs on \cite{Maker21e}. We opted for implementing our system entirely in Python, which consisted of rewriting the Solidity code from the Dai stablecoin system's Core Module, Collateral Module, Dai Module, Liquidation 2.0 Module, and part of its System Stabilizer Module (the \texttt{Vow}) — the components essential to get the system up and running. Beyond this, we created modules to represent keepers, abstractions for market conditions like collateral prices, gas prices, and liquidity, and an engine for running simulations and gathering statistics. See \textbf{Figure \ref{system}} for a diagram of the components we implemented (excluding the simulation engine), using the assets and styles defined in MakerDAO's system diagram \cite{Maker21f}.

We decided against running a local testchain with the Maker smart contracts themselves, for a number of reasons:

\begin{itemize}
    \item Complexity. Running a local testchain would require a lot of scaffolding, such as running a full node process, packaging transactions, and deploying all referenced smart contracts (even those we don't interact with), to mention a few. Operating at such a low level of abstraction would make some implementation shortcuts impossible (for example bypassing MakerDAO's Oracle Module), and would result in a deeply fragmented debugging process. Although great tools like Paradigm's Foundry \cite{Paradigm21} exist for working with a local testchain, we felt using them would not sufficiently wrangle the complexity.
    \item Speed. Even though we could tune block time on a local testchain to some extent, having to spin up, populate, and tear down a new chain for every simulation, along with all the processes that interact with it, would impose far too much compute and storage overhead for fast, iterative experimentation.
    \item Composability. Writing our smart contract modules, keepers, and simulation engine in Python allowed for efficient code reuse and high-level abstractions throughout our codebase. Additionally, it allowed for straightforward and granular passing of data between configuration files, keepers, smart contracts, the simulation engine, and the statistics and data visualization modules, without any marshalling concerns. Of course, this also made the use familiar Python packages for analysis quite seamless.
\end{itemize}

However, not running a local testchain made it quite difficult to simulate the gas costs of triggering liquidations, restarting auctions, and taking auctions. Thus, we looked towards historical gas costs \cite{Etherscan21} and chose to sample them from a normal distribution centered at 300000 gas.

\subsection{Experiment Structure}
We decided to simulate the protocol’s execution over historical data, as opposed to simulating market conditions themselves. Doing so would require very intricately defined mathematical processes such as geometric Brownian motion or Ornstein-Uhlenbeck \cite{PhysRev.36.823} to simulate collateral prices, and dynamic models of exchange liquidity and gas prices. Gauntlet’s report offers a glance at the complexity this introduces \cite{Gauntlet21b}. We felt that at our level of expertise, such complexity would lead to much less accurate results than the smaller sample space imposed by selecting from fitting historical timeframes.

We decided to focus only on ETH-collateralized vaults in our simulation, as it is a sufficiently volatile collateral to trigger liquidations and sufficiently popular in the MakerDAO protocol. It was fairly simple to find providers of exact historical ETH and gas prices. Finding liquidity data for the ETH/DAI market was much less straightforward, so we decided to use Uniswap's liquidity data as an estimate. Uniswap’s (at the time, V2) pricing model results in much higher slippage than centralized, orderbook exchanges \cite{Uniswap20}. This, combined with the fact that Uniswap accounts for a small subset of the total ETH/DAI market, makes it a great conservative estimate of liquidity.

To test the efficacy of various \texttt{chip} and \texttt{tip} settings, we wanted to target timeframes where many liquidations would be triggered, in order to get a good sampling of liquidated vault sizes and collateralization ratios. We also wanted to ensure that the incentive parameters result in a timely liquidation response to vaults becoming undercollateralized in cases when the system is at larger risk of insolvency, and be able to cleanly handle high liquidation throughput. The timeframes we ended up choosing were the following (each one day in length):

\begin{itemize}
    \item September 5\textsuperscript{th}, 2020
    \item January 11\textsuperscript{th}, 2021
    \item January 21\textsuperscript{st}, 2021
    \item February 22\textsuperscript{nd}, 2021
    \item May 19\textsuperscript{th}, 2021
    \item June 21\textsuperscript{st}, 2021
\end{itemize}

We selected \texttt{chip} and \texttt{tip} values that occupy a reasonable range based around the initial settings they were implemented with (\texttt{chip}: 0.08, \texttt{tip}: 1000). For each timeframe, we swept over all permutations of the following \texttt{chip} and \texttt{tip} values:

\begin{center}
\begin{tabular}{ |l|l| }
    \hline
    \textbf{Parameter} & \textbf{Values} \\
    \hline
    \texttt{chip} & 0.001, 0.01, 0.1 \\ 
    \hline
    \texttt{tip} & 100, 500, 1000 \\
    \hline
\end{tabular}
\end{center}

For every combination of timeframe and parameter setting, we ran 10 separate simulations.

\subsection{Simulation Lifecycle}
A simulation is instantiated by accepting a set of parameters including things like the number of each keeper type and their preferences (e.g. collateralization ratio, desired profit margin), constants within the smart contracts, statistics to track, number of timesteps in the simulation, and feeds to pull from, to name a few. Throughout its execution, the simulation keeps track of the internal state of all smart contracts, the keepers, the value of feeds (i.e. ETH price, gas price, and liquidity), and any defined statistics for the current timestep.

Within each timestep, each keeper generates “actions,” which are triggered in a random order (so as to emulate the variance in connectivity and latency that keepers may experience). Each action may have listeners attached to it that record the value of a defined statistic at the time of that action’s execution. Examples of actions include updating collateral prices, triggering liquidations, betting on collateral auctions, etc. - just to give an idea. Two special actions exist as well, which are not generated by any keepers: those marking the start and end of a timestep. These are necessarily “triggered” as the very first and very last actions in a timestep, but don't affect the simulation state. They are only used by statistic tracking functions, as for some statistics, it’s important that their value is reset, or calculated, at the beginning, or end, of a timestep.

Our timesteps represent 10-minute intervals, and as such, collateral prices within one of these timesteps represent the price at the start of that 10-minute interval in historic time. This abstracts away any price variance within the 10-minute interval, and also makes an implicit assumption that MakerDAO's oracles will update the system with a collateral’s price within 10 minutes of the price “occurring” on external exchanges. In actuality, this is deliberately not the case: MakerDAO's Oracle Security Module was built specifically to ensure that prices are only propagated through the system after a specified delay \cite{Maker21d}. Generally, our model abstracts away any mempool congestion and other network interaction beyond randomizing the execution of actions within a timestep. We felt this would not affect our simulation, as our keepers will always have enough time to react to price and prepare an according action before the price updates.

We tracked the following statistics in our analysis:

\begin{itemize}
    \item Number of liquidations per timestep
    \item Number of unsafe (undercollateralized) vaults per timestep
    \item Amount of incentives paid per timestep
    \item Average number of timesteps to liquidate unsafe vaults over the course of the simulation (“average response time”)
\end{itemize}

Using these statistics, we also composed a metric tracking, for a simulation with a given parameter setting over a given timeframe, the percent decrease in average response time per additional dollar of incentive, when comparing to a simulation over the same timeframe with the minimal parameter settings (\texttt{chip} = 0.001 and \texttt{tip} = 100).

\subsection{Keeper Models}
We created 5 different types of keepers, each with their own role to play in the system. An important simplification that we made was defining separate keeper models for kicking off liquidations or restarting auctions, and actually participating in auctions. Realistically, both types of actions are likely to be performed by the same keeper. In fact, it’s probable that a large part of their motivation to liquidate or restart is that they would know about the auction earlier than other keepers and have time to prepare their transactions for taking it. We deliberately made the decision to isolate these responsibilities in our simulation in order to consider optimal liquidation triggering and auction restarting in a vacuum, abstracting away the complexity of optimizing the auction lifecycle.

Our keeper models are defined below:

\begin{center}
\begin{tabular}{ |p{9em}|p{5em}|p{7em}| }
    \hline
    \textbf{Keeper type} & \textbf{Number of instances} & \textbf{Purpose} \\
    \hline
    \texttt{NaiveVaultKeeper} & 50 & Opens vaults \\ 
    \hline
    \texttt{SpotterKeeper} & 1 & Updates prices \\
    \hline
    \texttt{BarkKeeper} & 5 & Liquidates vaults \\
    \hline
    \texttt{RedoKeeper} & 5 & Restarts auctions \\
    \hline
    \texttt{NaiveClipperKeeper} & 5 & Bids on auctions \\
    \hline
\end{tabular}
\end{center}

\subsubsection{\texttt{NaiveVaultKeeper}}
The \texttt{NaiveVaultKeeper}'s sole function is opening the largest vault it can, in order to seed the system with vaults that may be liquidated over the course of the simulation. It’s defined by two main parameters: initial ETH collateral balance, and desired collateralization ratio. \texttt{NaiveVaultKeeper}s are instantiated with an initial balance that’s drawn from a normal distribution centered around 10 ETH, and a collateralization ratio drawn from a normal distribution centered around 175\%.

At the simulation's first timestep, the \texttt{NaiveVaultKeeper} opens a vault with all of its initial ETH collateral, drawing DAI up to its collateralization ratio, and then sits idle.

All other keepers inherit from the \texttt{NaiveVaultKeeper}’s functionality, in that all other keepers also open vaults.

\subsubsection{\texttt{SpotterKeeper}}
The \texttt{SpotterKeeper} is a very basic keeper that simply updates collateral prices. Every timestep, it pulls the current value from the collateral price feed, and pushes this value into the \texttt{Spotter} smart contract that is referenced by the rest of the Dai stablecoin system.

\subsubsection{\texttt{BarkKeeper}}
The \texttt{BarkKeeper} is one of two incentivized keepers. A \texttt{BarkKeeper} searches for undercollateralized vaults and liquidates them. Upon finding such an “unsafe” vault, the \texttt{BarkKeeper} will liquidate it if this is deemed profitable: namely, if the incentive payout to the \texttt{BarkKeeper} outweighs the gas costs of liquidation at current gas prices.

\subsubsection{\texttt{RedoKeeper}}
The \texttt{RedoKeeper} is the second incentivized keeper. It is very similar to the \texttt{BarkKeeper}, except that it restarts stale collateral auctions, as opposed to kicking off new ones.

Similarly to the \texttt{BarkKeeper}, the \texttt{RedoKeeper} will only restart an auction if it deems this to be profitable, where again profitability is determined by the incentive payout outweighing the gas costs of smart contract interaction at the current gas price.

\subsubsection{\texttt{NaiveClipperKeeper}}
The \texttt{NaiveClipperKeeper} is the simulation’s auction participant: it’s responsible for “taking” collateral auctions on liquidated vaults. Each timestep, every \texttt{NaiveClipperKeeper} iterates through all of the available auctions and takes them according to three conditions:

\begin{enumerate}
    \item The \texttt{NaiveClipperKeeper} has sufficient DAI capital to take the whole lot
    \item The bid price is lower than the desired discount of the \texttt{NaiveClipperKeeper}
    \item Taking the auction would be profitable
\end{enumerate}

The first condition is straightforward to understand, but note that it is not strictly necessary in Liquidations 2.0, as taking a fraction of the lot is now possible. Regarding the second condition, a \texttt{NaiveClipperKeeper}’s desired discount is its primary defining parameter, defining how far below market price the bid price of a collateral auction must be to fit the keeper's preference. As an example, a desired discount of 70\% means that a \texttt{NaiveClipperKeeper} would take a collateral auction when it’s bid price is 70\% of the collateral’s market price. The desired discount is drawn from a normal distribution centered around 85\%. Finally, in determining whether or not an auction settlement is deemed “profitable”, we assume that the \texttt{NaiveClipperKeeper} will immediately swap the collateral back to DAI on Uniswap. As such, taking an auction is profitable if the expected DAI proceeds from this swap (accounting for slippage cost through Uniswap) outweigh the gas costs of smart contract interaction (at the current gas price).

\section{Analysis}
We initially focused on the most bearish timeframe, namely that of May 19th, 2021, in order to find the parameter settings that properly service the timeframe with the highest throughput of liquidations. We began by simulating the minimal parameter settings of \texttt{chip}: 0.001 and \texttt{tip}: 100 to set a baseline to compare against.

\begin{figure}[t]
  \centering
  \includegraphics[width=\linewidth]{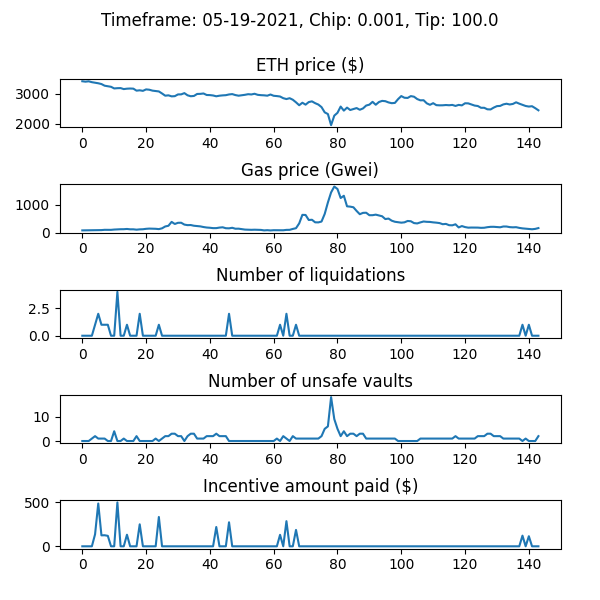}
  \caption{A sample run of the May 19th, 2021 timeframe with parameter settings \texttt{chip}: 0.001, \texttt{tip}: 100}
  \label{f1}
\end{figure}

Look to \textbf{Figure \ref{f1}} to see the development of our tracked statistics over one of the runs of this simulation. A healthy parameter setting would be evidenced by a minimal area under the curve for the "Number of unsafe vaults" graph. More concretely, we want this graph to be very "spiky," with its value dropping to 0 as quickly as possible, meaning that vaults are liquidated (and thus deleted) immediately after they become undercollateralized. This would also be evidenced by the "Number of liquidations" graph mirroring the shape of "Number of unsafe vaults" and lagging behind it minimally. Of course, we also want to minimize the area under the curve of the "Incentive amount paid" graph.

As expected, while this minimal parameter setting resulted in the least amount of incentives getting paid over the course of the simulation - an average of 4705.33 DAI across 10 separate runs - the average response time was incredibly poor, at 7.126 timesteps between a vault becoming undercollateralized and that vault being liquidated. Ideally, the average response time would be 1: liquidating vaults in the very next timestep after they’ve become unsafe (our simulation doesn’t allow for liquidating unsafe vaults in the same timestep).

\begin{figure}[t]
  \centering
  \includegraphics[width=\linewidth]{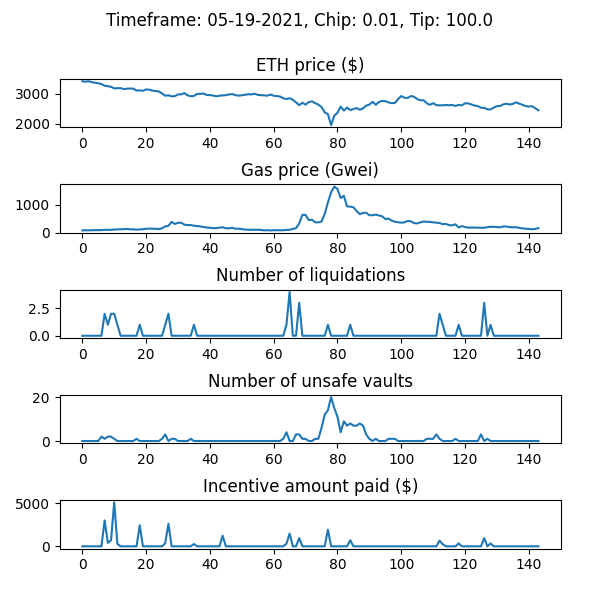}
  \caption{A sample run of the May 19th, 2021 timeframe with parameter settings \texttt{chip}: 0.01, \texttt{tip}: 100}
  \label{f2}
\end{figure}

Next, we looked towards the parameter setting of \texttt{chip}: 0.01, \texttt{tip}: 100, to see what effect increasing \texttt{chip} while leaving \texttt{tip} constant would have (see \textbf{Figure \ref{f2}}).

This already showed drastic improvement: an average response time of 1.053 timesteps, which is already incredibly close to the ideal response time of 1, and indicates a sufficiently healthy liquidation system. As expected, however, the average of the total incentive amount paid over the course of a simulation increased, approximately by a factor of 5, to 25201.93 DAI.

\begin{figure}[t]
  \centering
  \includegraphics[width=\linewidth]{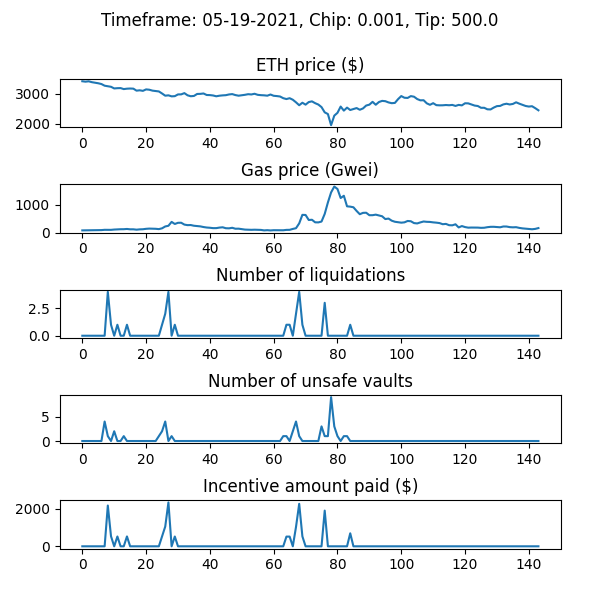}
  \caption{A sample run of the May 19th, 2021 timeframe with parameter settings \texttt{chip}: 0.001, \texttt{tip}: 500}
  \label{f3}
\end{figure}

Similarly, we examined the \texttt{chip}: 0.001, \texttt{tip}: 500 case to discern the effects of increasing \texttt{tip} while leaving \texttt{chip} constant (see \textbf{Figure \ref{f3}}).

Here, we came away with an average response time of 1.063 timesteps and an average incentive payout of 18436.78 DAI over the course of a simulation. Interestingly, the response time was slightly worse on average than the \texttt{chip}: 0.01, \texttt{tip}: 100 case, but there was a sizable reduction in average incentive payout.

To capture the efficacy of a given parameter setting, we used our metric of average percent decrease in response time per additional dollar of incentive, relative to the \texttt{chip}: 0.001, \texttt{tip}: 100 baseline. In the \texttt{chip}: 0.01, \texttt{tip}: 100 case, we have, on average, a 0.00416\% decrease in response time per additional dollar of incentive, and in the \texttt{chip}: 0.001, \texttt{tip}: 500 case, we have an average 0.0062\% decrease. This means that increasing the \texttt{tip} to 500 while leaving \texttt{chip} at a minimum was approximately 1.5x more cost-effective than increasing \texttt{chip} to 0.01 while leaving \texttt{tip} at a minimum.

\begin{figure}[t]
  \centering
  \includegraphics[width=\linewidth]{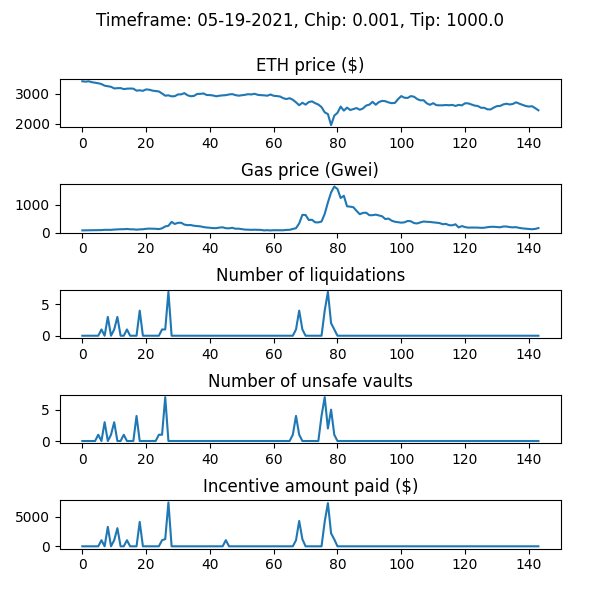}
  \caption{A sample run of the May 19th, 2021 timeframe with parameter settings \texttt{chip}: 0.001, \texttt{tip}: 1000}
  \label{f4}
\end{figure}

In an effort to extrapolate this trend and approach closer to the ideal case of immediate response time (1 timestep), we analyzed the performance of the \texttt{chip}: 0.001, \texttt{tip}: 1000 case (see \textbf{Figure \ref{f4}}).

We did, in fact, arrive at an average response time of 1 timestep across all liquidated vaults in all simulations. However, the average incentive payout was 38725.42 DAI over the course of a simulation. This bore an average decrease in response time of 0.00253\% per additional dollar of incentive, making it approximately 0.4x as cost-effective as the \texttt{chip}: 0.001, \texttt{tip}: 500 case above.

The \texttt{chip}: 0.001, \texttt{tip}: 500 case resulted in comfortably fast average response times across the remaining timeframes, ranging from 1.078 to 1.402 and averaging at 1.155. In fact, we found that even \texttt{chip}: 0.001, \texttt{tip}: 100, worked fairly well across the other timeframes, bearing response times that ranged from 1.085 to 2.654 and averaging at 1.415. However, given this parameter setting's particularly poor average response time in the May 19th, 2021 timeframe, we still advise for prudency and err on the side of the \texttt{chip}: 0.001, \texttt{tip}: 500 parameter setting.

\section{Conclusion}
When we began running simulations in summer of 2021, the \texttt{chip} and \texttt{tip} parameters on the ETH-A \texttt{Clipper} contract were initially set to 0.08 and 1000, respectively. On October 22nd, 2021, they were set to 0.001 and 300, as suggested by the research presented in a risk assessment for onboarding stETH as a collateral type \cite{Lido21}. At the time of writing, these parameters are still active, and are extremely consistent with our own, entirely independent, research findings. In short, we would not recommend changing the \texttt{chip} and \texttt{tip} parameters.

However, keeping in mind the assumptions and simplifications made throughout our research, we do not strongly endorse our exact parameter settings of \texttt{chip}: 0.001, \texttt{tip}: 500. What we do, however, stand by, is the higher-level takeaway that it is more cost-effective to increase the \texttt{tip} parameter than the \texttt{chip} parameter.

We are particularly in favor of this stance because of the Maker protocol's standing problem of small vaults not being deemed worth liquidating by keepers. Previously, the approach to this issue was to steadily increase the value of the \texttt{dust} parameter denoting the minimum size of a vault \cite{Maker21g}, forcing users to open vaults large enough worth liquidating, should this be necessary. Of course, this makes the system less and less accessible to more capital-constrained users. Instead, providing sufficient liquidation incentives that allow users to open small vaults is a much more attractive option. To this end, relying more heavily on a flat-rate incentive (\texttt{tip}) as opposed to one that is proportional to the size of the vault (\texttt{chip}) is a much more sensible solution.

Looking forward, we highly encourage that interested readers extend our system to test any other aspects of Liquidations 2.0 incentives, or of the Maker protocol in general. In particular, it would be important to ensure that incentives are balanced to protect against the incentive farming attack laid out in the Liquidations 2.0 documentation \cite{Maker21h}.

For the purposes of general experimentation with the Maker protocol, we have also implemented the Liquidations 1.2 Module (the \texttt{Flipper}), the remaining components of the System Stabilizer Module (the debt and surplus auctions, \texttt{Flop} and \texttt{Flap}), and some basic liquidation and auction keepers for Liquidations 1.2.

We welcome any contributions to, or experimentation with, our codebase: \url{https://github.com/BerkeleyBlockchain/stablesims} \cite{B@B21}.

\bibliographystyle{ACM-Reference-Format}
\bibliography{main}


\begin{thebibliography}{18}


\ifx \showCODEN    \undefined \def \showCODEN     #1{\unskip}     \fi
\ifx \showDOI      \undefined \def \showDOI       #1{#1}\fi
\ifx \showISBNx    \undefined \def \showISBNx     #1{\unskip}     \fi
\ifx \showISBNxiii \undefined \def \showISBNxiii  #1{\unskip}     \fi
\ifx \showISSN     \undefined \def \showISSN      #1{\unskip}     \fi
\ifx \showLCCN     \undefined \def \showLCCN      #1{\unskip}     \fi
\ifx \shownote     \undefined \def \shownote      #1{#1}          \fi
\ifx \showarticletitle \undefined \def \showarticletitle #1{#1}   \fi
\ifx \showURL      \undefined \def \showURL       {\relax}        \fi
\providecommand\bibfield[2]{#2}
\providecommand\bibinfo[2]{#2}
\providecommand\natexlab[1]{#1}
\providecommand\showeprint[2][]{arXiv:#2}

\bibitem[\protect\citeauthoryear{Adams, Zinmeister, and Robinson}{Adams
  et~al\mbox{.}}{2020}]%
        {Uniswap20}
\bibfield{author}{\bibinfo{person}{H. Adams}, \bibinfo{person}{N. Zinmeister},
  {and} \bibinfo{person}{D. Robinson}.} \bibinfo{year}{2020}\natexlab{}.
\newblock \bibinfo{title}{Uniswap v2 Core}.
\newblock
\newblock
\urldef\tempurl%
\url{https://uniswap.org/whitepaper.pdf}
\showURL{%
\tempurl}


\bibitem[\protect\citeauthoryear{at~Berkeley}{at~Berkeley}{2021}]%
        {B@B21}
\bibfield{author}{\bibinfo{person}{Blockchain at Berkeley}.}
  \bibinfo{year}{2021}\natexlab{}.
\newblock \bibinfo{title}{StableSims}.
\newblock
\newblock
\urldef\tempurl%
\url{https://github.com/BerkeleyBlockchain/stablesims}
\showURL{%
\tempurl}


\bibitem[\protect\citeauthoryear{DAO}{DAO}{2021}]%
        {Lido21}
\bibfield{author}{\bibinfo{person}{Lido DAO}.} \bibinfo{year}{2021}\natexlab{}.
\newblock \bibinfo{title}{[stETH] Collateral Onboarding Risk Evaluation}.
\newblock
\newblock
\urldef\tempurl%
\url{https://forum.makerdao.com/t/steth-collateral-onboarding-risk-evaluation/9061}
\showURL{%
\tempurl}


\bibitem[\protect\citeauthoryear{Etherscan}{Etherscan}{2021}]%
        {Etherscan21}
\bibfield{author}{\bibinfo{person}{Etherscan}.}
  \bibinfo{year}{2021}\natexlab{}.
\newblock \bibinfo{title}{ETH-A Clipper Contract}.
\newblock
\newblock
\urldef\tempurl%
\url{https://etherscan.io/address/0xc67963a226eddd77B91aD8c421630A1b0AdFF270}
\showURL{%
\tempurl}


\bibitem[\protect\citeauthoryear{Foundation}{Foundation}{2020a}]%
        {Maker20a}
\bibfield{author}{\bibinfo{person}{Maker Foundation}.}
  \bibinfo{year}{2020}\natexlab{a}.
\newblock \bibinfo{title}{A Liquidation System Redesign: A Pre-MIP Discussion}.
\newblock
\newblock
\urldef\tempurl%
\url{https://forum.makerdao.com/t/a-liquidation-system-redesign-a-pre-mip-discussion/2790}
\showURL{%
\tempurl}


\bibitem[\protect\citeauthoryear{Foundation}{Foundation}{2020b}]%
        {Maker20b}
\bibfield{author}{\bibinfo{person}{Maker Foundation}.}
  \bibinfo{year}{2020}\natexlab{b}.
\newblock \bibinfo{title}{The Market Collapse of March 12-13, 2020: How It
  Impacted MakerDAO}.
\newblock
\newblock
\urldef\tempurl%
\url{https://blog.makerdao.com/the-market-collapse-of-march-12-2020-how-it-impacted-makerdao/}
\showURL{%
\tempurl}


\bibitem[\protect\citeauthoryear{Foundation}{Foundation}{2021a}]%
        {Maker21g}
\bibfield{author}{\bibinfo{person}{Maker Foundation}.}
  \bibinfo{year}{2021}\natexlab{a}.
\newblock \bibinfo{title}{Glossary (Vat - Vault Engine)}.
\newblock
\newblock
\urldef\tempurl%
\url{https://docs.makerdao.com/smart-contract-modules/core-module/vat-detailed-documentation#glossary-vat-vault-engine}
\showURL{%
\tempurl}


\bibitem[\protect\citeauthoryear{Foundation}{Foundation}{2021b}]%
        {Maker21h}
\bibfield{author}{\bibinfo{person}{Maker Foundation}.}
  \bibinfo{year}{2021}\natexlab{b}.
\newblock \bibinfo{title}{Incentive Farming}.
\newblock
\newblock
\urldef\tempurl%
\url{https://docs.makerdao.com/smart-contract-modules/dog-and-clipper-detailed-documentation#incentive-farming}
\showURL{%
\tempurl}


\bibitem[\protect\citeauthoryear{Foundation}{Foundation}{2021c}]%
        {Maker21b}
\bibfield{author}{\bibinfo{person}{Maker Foundation}.}
  \bibinfo{year}{2021}\natexlab{c}.
\newblock \bibinfo{title}{Liquidation 2.0 Module}.
\newblock
\newblock
\urldef\tempurl%
\url{https://docs.makerdao.com/smart-contract-modules/dog-and-clipper-detailed-documentation}
\showURL{%
\tempurl}


\bibitem[\protect\citeauthoryear{Foundation}{Foundation}{2021d}]%
        {Maker21a}
\bibfield{author}{\bibinfo{person}{Maker Foundation}.}
  \bibinfo{year}{2021}\natexlab{d}.
\newblock \bibinfo{title}{The Maker Protocol: MakerDAO's Multi-Collateral Dai
  (MCD) System}.
\newblock
\newblock
\urldef\tempurl%
\url{https://makerdao.com/whitepaper/}
\showURL{%
\tempurl}


\bibitem[\protect\citeauthoryear{Foundation}{Foundation}{2021e}]%
        {Maker21f}
\bibfield{author}{\bibinfo{person}{Maker Foundation}.}
  \bibinfo{year}{2021}\natexlab{e}.
\newblock \bibinfo{title}{The Maker Protocol Smart Contract Modules System}.
\newblock
\newblock
\urldef\tempurl%
\url{https://docs.makerdao.com/#the-maker-protocol-smart-contract-modules-system}
\showURL{%
\tempurl}


\bibitem[\protect\citeauthoryear{Foundation}{Foundation}{2021f}]%
        {Maker21c}
\bibfield{author}{\bibinfo{person}{Maker Foundation}.}
  \bibinfo{year}{2021}\natexlab{f}.
\newblock \bibinfo{title}{MIP45: Liquidations 2.0 (LIQ-2.0) - Liquidation
  System Redesign}.
\newblock
\newblock
\urldef\tempurl%
\url{https://forum.makerdao.com/t/mip45-liquidations-2-0-liq-2-0-liquidation-system-redesign/6352}
\showURL{%
\tempurl}


\bibitem[\protect\citeauthoryear{Foundation}{Foundation}{2021g}]%
        {Maker21e}
\bibfield{author}{\bibinfo{person}{Maker Foundation}.}
  \bibinfo{year}{2021}\natexlab{g}.
\newblock \bibinfo{title}{Multi Collateral Dai}.
\newblock
\newblock
\urldef\tempurl%
\url{https://github.com/makerdao/dss}
\showURL{%
\tempurl}


\bibitem[\protect\citeauthoryear{Foundation}{Foundation}{2021h}]%
        {Maker21d}
\bibfield{author}{\bibinfo{person}{Maker Foundation}.}
  \bibinfo{year}{2021}\natexlab{h}.
\newblock \bibinfo{title}{Oracle Security Module (OSM) - Detailed
  Documentation}.
\newblock
\newblock
\urldef\tempurl%
\url{https://docs.makerdao.com/smart-contract-modules/oracle-module/oracle-security-module-osm-detailed-documentation}
\showURL{%
\tempurl}


\bibitem[\protect\citeauthoryear{Networks}{Networks}{2021a}]%
        {Gauntlet21a}
\bibfield{author}{\bibinfo{person}{Gauntlet Networks}.}
  \bibinfo{year}{2021}\natexlab{a}.
\newblock \bibinfo{title}{MakerDAO Auction Assessment}.
\newblock
\newblock
\urldef\tempurl%
\url{https://maker-report.gauntlet.network/}
\showURL{%
\tempurl}


\bibitem[\protect\citeauthoryear{Networks}{Networks}{2021b}]%
        {Gauntlet21b}
\bibfield{author}{\bibinfo{person}{Gauntlet Networks}.}
  \bibinfo{year}{2021}\natexlab{b}.
\newblock \bibinfo{title}{MakerDAO Auction Assessment: Model Definition}.
\newblock
\newblock
\urldef\tempurl%
\url{https://maker-report.gauntlet.network/model}
\showURL{%
\tempurl}


\bibitem[\protect\citeauthoryear{Paradigm}{Paradigm}{2021}]%
        {Paradigm21}
\bibfield{author}{\bibinfo{person}{Paradigm}.} \bibinfo{year}{2021}\natexlab{}.
\newblock \bibinfo{title}{Foundry}.
\newblock
\newblock
\urldef\tempurl%
\url{https://github.com/gakonst/foundry}
\showURL{%
\tempurl}


\bibitem[\protect\citeauthoryear{Uhlenbeck and Ornstein}{Uhlenbeck and
  Ornstein}{1930}]%
        {PhysRev.36.823}
\bibfield{author}{\bibinfo{person}{G.~E. Uhlenbeck} {and}
  \bibinfo{person}{L.~S. Ornstein}.} \bibinfo{year}{1930}\natexlab{}.
\newblock \showarticletitle{On the Theory of the Brownian Motion}.
\newblock \bibinfo{journal}{\emph{Phys. Rev.}}  \bibinfo{volume}{36}
  (\bibinfo{date}{Sep} \bibinfo{year}{1930}), \bibinfo{pages}{823--841}.
\newblock
Issue 5.
\urldef\tempurl%
\url{https://doi.org/10.1103/PhysRev.36.823}
\showDOI{\tempurl}


\end{thebibliography}

\end{document}